\documentclass[twocolumn,superscriptaddress,aps,pre,bibtex]{revtex4-1}
\usepackage[T1]{fontenc}
\usepackage[latin9]{inputenc}

\usepackage{color}
\usepackage{nicefrac}
\usepackage{float}
\usepackage{amsmath}
\usepackage{amssymb}
\usepackage{graphicx}
\usepackage{subfigure}

\newcommand{\newb}[1]{{\color{black}#1}}
\begin{document}

\title{Timeseries thresholding and the definition of avalanche size}

\author{Pablo Villegas} \affiliation{Departamento de Electromagnetismo
  y F{\'\i}sica de la Materia e Instituto Carlos I de F{\'\i}sica
  Te\'orica y Computacional. Universidad de Granada.  E-18071,
  Granada, Spain} 
\author{Serena di Santo} \affiliation{Departamento de
  Electromagnetismo y F{\'\i}sica de la Materia e Instituto Carlos I
  de F{\'\i}sica Te\'orica y Computacional. Universidad de Granada.
  E-18071, Granada, Spain} \affiliation{Dipartimento di Scienze
  Matematiche, Fisiche e Informatiche, Universit\`a di Parma, via
  G.P. Usberti, 7/A - 43124, Parma, Italy} \affiliation{INFN, Gruppo
  Collegato di Parma, via G.P. Usberti, 7/A - 43124, Parma, Italy}
\author{Raffaella Burioni} \affiliation{Dipartimento
  di Scienze Matematiche, Fisiche e Informatiche, Universit\`a di
  Parma, via G.P. Usberti, 7/A - 43124, Parma, Italy}
\affiliation{INFN, Gruppo Collegato di Parma, via G.P. Usberti, 7/A -
  43124, Parma, Italy} \author{Miguel A. Mu\~noz}
\affiliation{Departamento de Electromagnetismo y F{\'\i}sica de la
  Materia e Instituto Carlos I de F{\'\i}sica Te\'orica y
  Computacional. Universidad de Granada.  E-18071, Granada,
  Spain}\affiliation{Dipartimento di Scienze Matematiche, Fisiche e
  Informatiche, Universit\`a di Parma, via G.P. Usberti, 7/A - 43124,
  Parma, Italy}
\begin{abstract}
  \newb{Avalanches whose sizes and durations are distributed as power
    laws appear in many contexts, from physics to geophysics and
    biology.  } Here, we show that there is a hidden peril in
  thresholding continuous times series --either from empirical or
  synthetic data-- for the identification of avalanches. In
  particular, we consider two possible alternative definitions of
  avalanche size used e.g. in the empirical determination of avalanche
  exponents in the analysis of neural-activity data.  By performing
  analytical and computational studies of an Ornstein-Uhlenbeck
  process (taken as a guiding example) we show that (i) if relatively
  large threshold values are employed to determine the beginning and
  ending of avalanches and (ii) if --as sometimes done in the
  literature-- avalanche sizes are defined as the total area (above
  zero) of the avalanche, then true asymptotic scaling behavior is not
  seen, instead the observations are dominated by transient effects.
  \newb{This problem --that we have detected in some recent works--
    leads to misinterpretations of the resulting scaling regimes.  }
\end{abstract}
\maketitle

\section{Introduction}

Episodic outbursts of activity or ``avalanches'' of highly variable
durations and sizes are observed in a large variety of scenarios in
condensed matter physics (vortices of type II superconductors
\cite{Altshuler} and Barkhaussen noise \cite{ABBM,Zapperi}),
high-energy astrophysics (X-ray flares \cite{Wang2013}), geophysics
(earthquakes \cite{Bak2002}), meteorology (rainfall \cite{Peters}),
neuroscience (neuronal avalanches \cite{BP2003}), as well as in other
biological systems (gene knock-out cascades) \cite{Serra-review}) and
man-made systems (failures on electrical power grids
\cite{powergrids}).  The probability distributions of sizes and
durations of such avalanches often exhibit a ``fat-tail'' that can be
fitted as a power-law distribution; i.e. the fingerprint of scaling
behavior.  Such scaling or scale invariance is often considered as
evidence of underlying criticality and many of the above systems are
claimed to operate at (tuned or self-organized) critical points
\cite{Bak,BTW,Jensen,Pruessner,BJP}.  In particular, in the context of
biology the idea that living systems (parts, aspects or groups of
them) may extract important functional advantages from operating at
criticality --i.e. at the edge of two different phases-- has been
deeply explored in recent years \cite{Mora-Bialek,MAM-rmp}.
 
In this regard, groundbreaking experimental evidence by Beggs and
Plenz \cite{BP2003}, revealed the existence of scale-invariant
episodes of electrochemical activity in neural tissues thereafter
named \emph{neural avalanches}.  Subsequently, neural avalanches were
robustly detected across a large variety of experimental settings,
tissues and species
\cite{BP2003,Plenz2007,Beggs2008,Arcangelis2012,Petermann2009,Hahn2010,Palva2012,Plenz2015-pyramidal}.
In particular, neuronal avalanche sizes, $S$, were robustly observed
to be distributed as a power-law $P(S) \sim S^{-\tau}$ with
$\tau \approx 3/2$ up to some upper cut-off; similarly, avalanche
durations $T$ were well fitted by $P(T)\sim T^{-\alpha}$ with
$\alpha \approx 2$ up to some characteristic maximum time
\cite{BP2003}. Furthermore, fundamental scaling relationships
\cite{Avalanches} were observed to be fulfilled: e.g.  the averaged
avalanche size scales as $\langle S\rangle \sim T^\gamma$ and the set
of exponents obey $\gamma=(\alpha-1)/(\tau-1)$ \cite{Beggs2012}.

This set of empirically reported exponent values is in agreement with
that of the well-known critical (or ``unbiased'') branching processes,
also called Galton-Watson process, originally introduced to describe
the statistics of the extinction of family names)
\cite{Watson,Liggett,SOBP,Redner}. Actually, the set of exponent
values $\tau=3/2$, $\alpha=2$ and $\gamma=2$ are extremely universal
as they are shared by many different propagation processes in
high-dimensional systems as well as in many types of networks
\cite{ABBM,Newman}.  In particular, they are the mean-field exponents
shared by models such as the contact process, directed and isotropic
percolation, susceptible-infected-susceptible, and a large list of
other prototypical models for spreading/propagation dynamics above
their respective upper critical dimensions
\cite{Avalanches,Binney,Wiese,Corral,JABO2}.

Thus, it was conjectured that neuronal systems might operate close to
the edge of marginal propagation of neural (electro-chemical) activity
\cite{BP2003,Chialvo2004}, opening the door to exciting theoretical
perspectives and some debate (see \cite{MAM-rmp} for a recent review).
However, as extensively discussed in the literature, diverse
  generative processes for the emergence of power-laws exist
  \cite{Critics,Newman-powerlaws,Mitzenmacher2004}, and not all
  power-law distributions can be taken as a signature of criticality.
  For instance, a diverging correlation length needs to be identified
  to assign a given phenomenon to criticality. In recent years,
  some authors  have suggested that the origin of the observed
  power-law scaling in neural systems might stem from other types of
  criticality (rather than marginal propagation)
  \cite{neutral-neural,PNAS-LG,SOB} or even be unrelated to critical
  behavior \cite{Touboul,Viola-Shriki}.

  In the present brief paper --leaving aside the putative connection
  with criticality-- we contribute with an additional piece of
  information to the already controversial discussion about the
  statistics of neuronal avalanches. In particular, we show that some
  of the reported empirical evidence in favor of the value $\tau=3/2$
  --and thus, seemingly in favor of the existence of an underlying
  critical branching process-- might be misleading as there is a
  technical problem in the way avalanches are measured, which hinders
  the observation of the true asymptotic behavior.  More in general,
  we underline that particular attention needs to be taken when
  avalanches of activity --defined by thresholding-- are inferred from
  a continuous time series of activity. Our findings, add to the
    recent literature warning on the ``perils'' associated with
    thresholding in timeseries \footnote{For instance, since the same
    avalanche could e.g. be split in two by the effect of a higher
    threshold, it would introduce correlations between different
    avalanches, altering the measurement of exponents
    \cite{Perils1,Perils2}.}  \cite{Perils1,Perils2}.

\section{Definition of avalanches}

\newb{As discussed in the introduction, avalanching phenomena are best
  described, at least in mean-field, as branching processes.  Such
  processes can reach the value $0$ which is an absorbing state:
  avalanches are naturally defined as excursions away from such a
  state caused by small perturbations \cite{Branching}.  However, in
  many contexts --including neuroscience but not only-- the term
  ``avalanche'' is used to refer to excursions of time series above
  some given (arbitrary) threshold, regardless of absorbing states the
  existence of any absorbing state.  In this section we discuss such
  avalanches and their statistics.}

\subsubsection{Avalanches in the Wiener and Ornstein-Uhlenbeck processes}
Let us consider, for argument's sake, a time series for a stochastic
real variable $x$ --as illustrated in Fig.\ref{fig:OU}-- generated by
a Wiener Process, i.e. by a continuous-time unbiased random walk (RW)
defined by the following Langevin equation \cite{Gardiner}:
\begin{equation}
\dot{x}(t)=\sigma\eta(t),
\end{equation}
where $\eta(t)$ is a Gaussian white noise with zero mean and unit
variance, and $\sigma$ is the noise amplitude.  For such a time series
(which can be thought as describing the time course of the activity of
some arbitrary system) the duration $T$ of an avalanche is the amount
of time for which $x$ stays above a given threshold, i.e. an avalanche
begins/ends when the activity signal crosses the threshold from
below/above; the avalanche size $S$ is the area covered between the
walk trajectory and the threshold reference line (see Fig. \ref{fig:OU}).
Observe that similarly, given the symmetry of the process, one could
also define avalanches as excursions below threshold.

The probability distribution of avalanche durations $T$ can be
straightforwardly identified with the first-return time statistics of
random walks (see Fig.1) which is well-known to scale with an exponent
$\alpha=3/2$. Similarly, also the size-distribution exponent
$\tau=4/3$ and the remaining exponent $\gamma=3/2$ are well-known for
random walks (see Fig. \ref{fig:OU} and Table I); pedagogical 
derivations of these results, as well as a comparison with the
branching process class can be found in
e.g. \cite{Branching,Artime}.
Importantly, these results for the random walk do not depend of the
value of the chosen threshold.
\begin{table}[H]
\begin{center}
\begin{tabular}{|c|c|c|c|}
\hline
 & $P(S)\sim S^{-\tau}$  & $P(T)\sim S^{-\alpha}$  & $P(S\mid T)\sim T^{\gamma}$\tabularnewline
\hline 
BP  & $\tau=3/2$  & $\alpha=2$  & $\gamma=2$\tabularnewline
\hline 
RW  & $\tau=4/3$  & $\alpha=3/2$  & $\gamma=3/2$\tabularnewline
\hline
\end{tabular}
\caption{Summary of the avalanche (mean-field) exponents: size
  $(\tau)$, duration $(\alpha)$ and averaged avalanche size $(\gamma)$
  for the (un-biased) branching process (BP) and the (un-biased)
  random walk (RW); see e.g. \cite{Branching}.}
\end{center}
\label{table:exp}
\end{table}
\begin{center}
\begin{figure}[H]
\begin{centering}
\includegraphics[width=1\columnwidth]{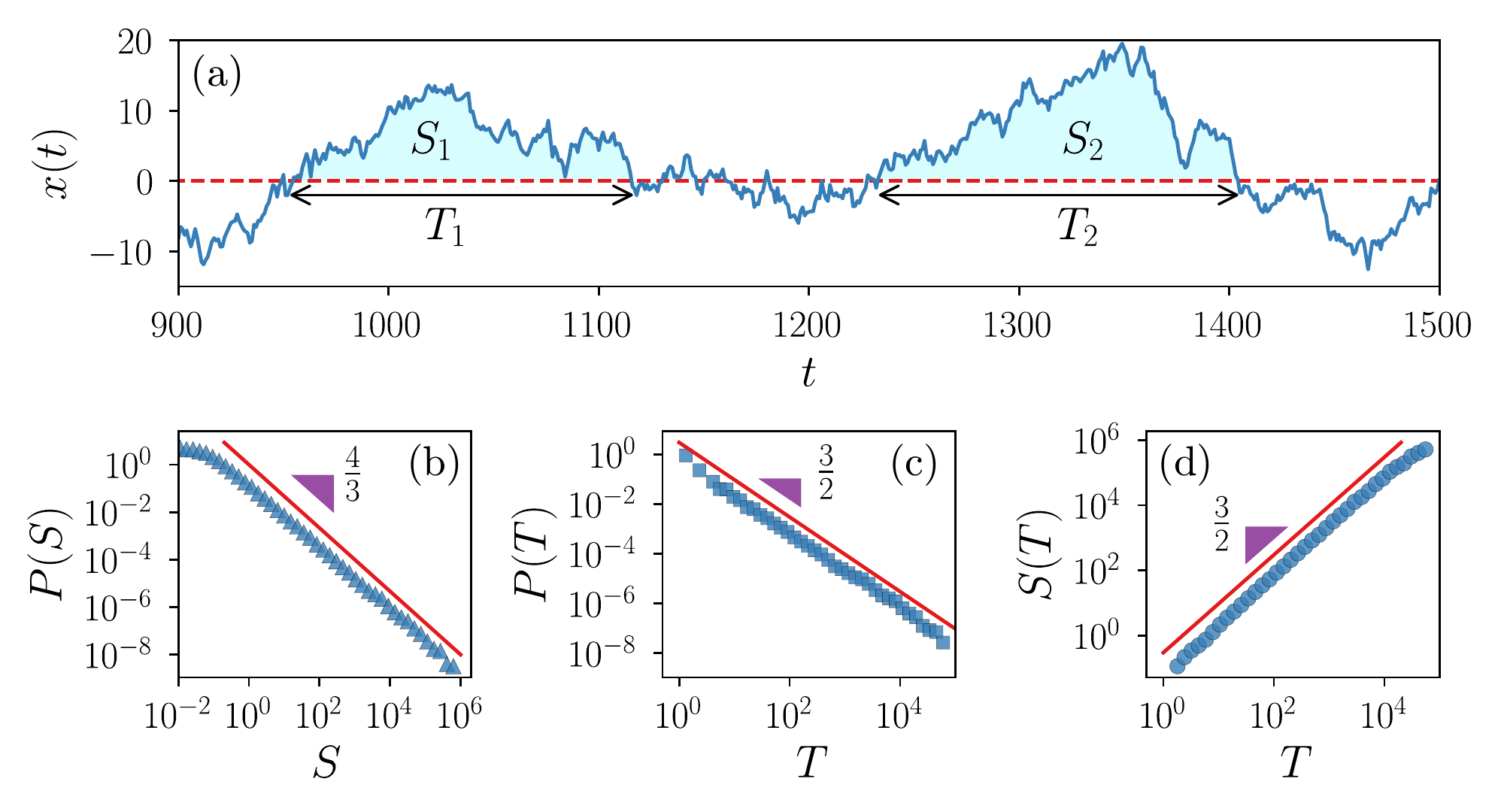}
\par\end{centering}
\caption[Illustration of the first return time statistics of a Random
Walk]{(Color online) Illustration of how avalanche duration $T$ and
  size $S$ are defined for an unbiased random walk.  (a)
  Illustration of a particular time series, in which two avalanches of
  durations $T_1$ and $T_2$ and sizes, $S_1$ and $S_2$, respectively,
  are emphasized.  The threshold is set to $0$ in this case (red
  dashed line). Lower panels show the probability
  distributions of: (b) sizes, (c) durations, and (d) average size for a fixed
  given duration (straight lines correspond to the well-known
  analytical predictions while symbols stand for computational
  results).
  \label{fig:OU} }
\end{figure}
\par\end{center}

In the more general case in which the walker is confined to hover
around a given mean value, one can describe the problem, in first
approximation, as an Ornstein-Uhlenbeck process \cite{Gardiner}:
\begin{equation}
\dot{x}(t)=-ax(t)+\sigma\eta(t),
\end{equation}
where there is an additional linear force term, $-ax$ (corresponding
to the negative derivative of the parabolic potential bounding the
walker close to $x=0$). Such a force introduces an upper cutoff in the
first-return times statistics of the unbiased RW (see
e.g. \cite{Artime} for a detailed derivation).  Thus, avalanches
intended as excursions above a given threshold in a process with a
well-defined steady-state value, have power-law-distributed sizes and
durations, with the exponents of the RW class (as in table I and as in
Fig.1), but only up to an upper-cut-off scale controlled by $1/a$,
such that it goes to infinity, i.e. it disappears when $a$ vanishes.

As a corollary of all this, let us remark that many real time series
describing (e.g. biological) problems in which some stochastic
variable fluctuates \emph{symmetrically} around a given mean value
exhibit effective avalanching behavior that --up to certain scale of
size and time-- can be described by the exponent values of the RW.
Let us stress again that, as argued above, it can be a matter of
  debated whether this type of behavior --describable in terms of
  random-walk excursions above a threshold-- can be called
  ``avalanching''. Actually, for most of the examples of interest in
  physics, as discussed in the first paragraph of the Introduction,
  this does not constitute an adequate description as it does not
  include any absorbing state.

\subsubsection{On the definition of avalanche size}
In many other circumstances time series exhibit \emph{asymmetric }
excursions around a mean value and/or may become trapped at some
absorbing state. An example of this is obtained when the variable
under scrutiny is a positive-definite density (e.g. of neural
activity), which, by definition, is constrained to take positive
values, $x(t) >0$.  In such cases, especially in the ones when there
is always some lingering activity so that the zero-value is hardly
reached (see Fig.2) a threshold $\theta>0$ is often employed to define
avalanches as periods during which the activity remains above such a
threshold.  In these situations, two alternative possibilities are
often used in the literature to measure the size of a so-defined
avalanches \cite{Dalla2018}:

(A) Following the random walk analogy, as done above, for a given
avalanche, one can define its size $S$ as the area in between the
time-series curve and the threshold ($\theta$) reference line.

(B) Alternatively, one can define the avalanche size $\Sigma$ as the
\emph{overall} integral of the time series during the avalanche,
i.e. above the reference line $x=0$ (see
e.g. \cite{Linken2012,Larremore2014} but there are other works making
this choice).
\begin{center}
\begin{figure}[h]
\begin{centering}
\includegraphics[width=1\columnwidth]{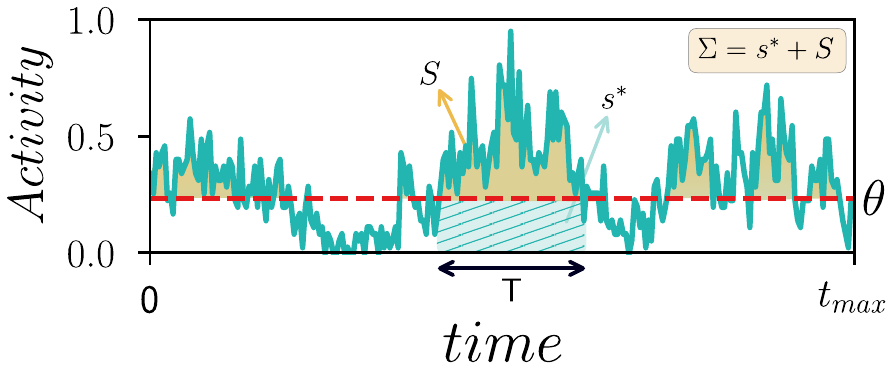}
\par\end{centering}
\caption[Analysis of first-passage times in a stochastic
process]{(Color online) Sketch of a non-symmetric stochastic process
  for a positive definite variable (describing, e.g. density of neural
  activity). $\theta$ (red dashed line) signals the arbitrarily fixed
  threshold employed to define avalanches.  For the large avalanche in
  the center of the graph, $S$ is the avalanche size using criterion A
  (area above threshold, colored in orange) and $T$ is its duration.
  On the other hand, using criterion B, $\Sigma=S+s^*$ (where $s^*$ is
  the area of the rectangle between zero and the threshold, colored in
  blueish color, with $s^*\propto T$) is an often-used alternative
  definition of avalanche size. As discussed in the text this
  definition may induce misleading interpretations of the resulting
  exponents.
  \label{fig:poil}}
\end{figure}
\par\end{center}
The difference between the two criteria to define avalanche sizes
  is sketched in Figure \ref{fig:poil}.  $\Sigma$, the total integral
of the activity is equal to $\Sigma=s^*+S$, where $S$ is the integral
of the signal above threshold, and $s^*$ is the area of the rectangle
under the threshold.

In what follows, we compare the statistics of avalanches obtained
using these two alternative definitions of size A and B for an
Ornstein-Uhlenbeck process. This will serve as an illustration of 
a  more general phenomenon that may also occur for other processes,
such as the one sketched in Fig.2.
\begin{center}
\begin{figure}[h]
\begin{centering}
\includegraphics[width=1\columnwidth]{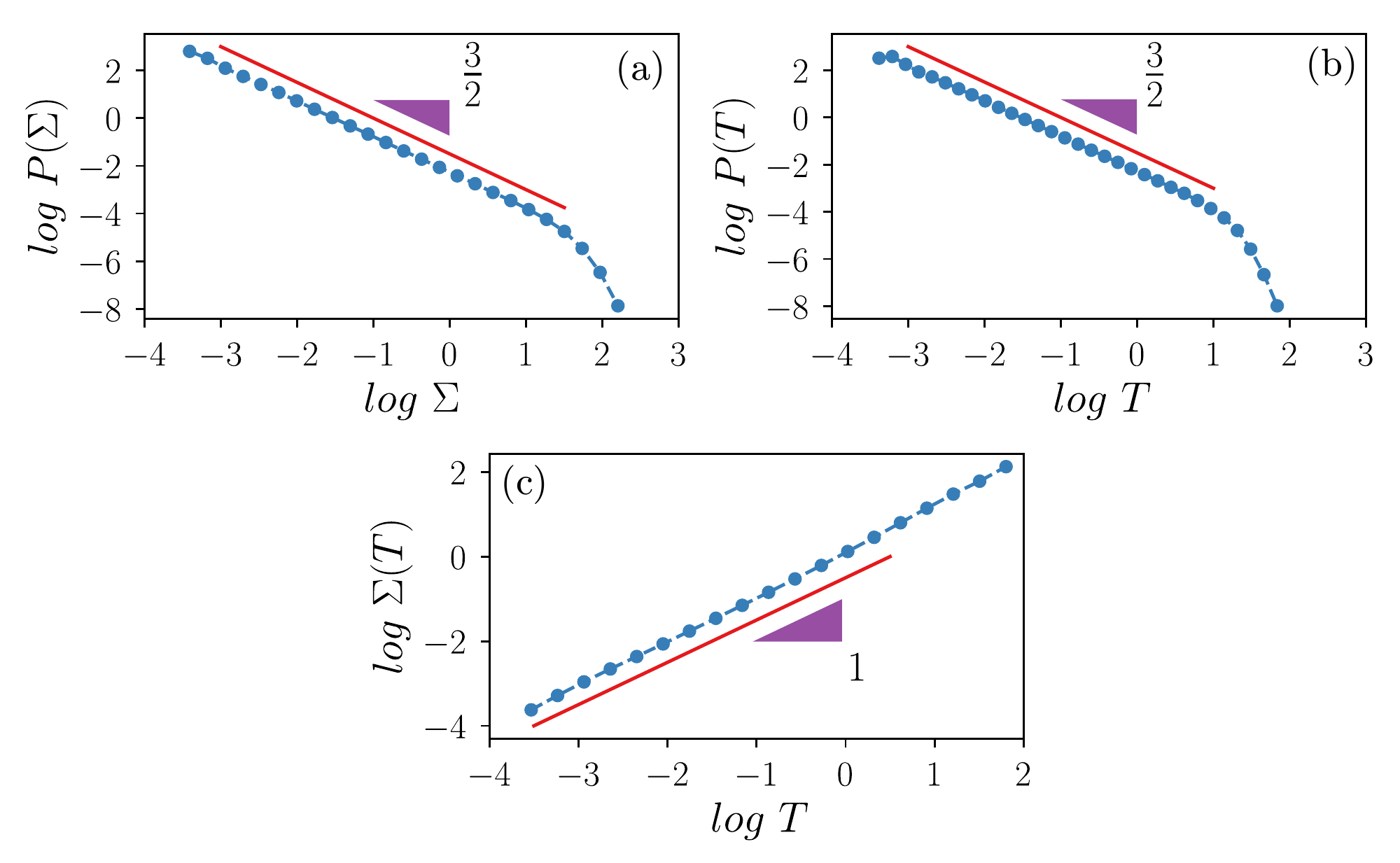}
\par\end{centering}
\caption[Power law distributions of the misleading avalanches]{(Color
  online) Statistics of avalanches of activity in a stochastic process
  (Ornstein-Uhlenbeck with $a=0.1$) employing $\Sigma$ as a
  measure of the avalanche size.  Observe that both, (a)
  avalanche-size and (b) avalanche-duration distributions obey scaling
  with the same exponent value for many orders of magnitude.  However,
  the exponent values $\tau=\alpha =3/2 $ and, consequently, as depicted in (c),
  $\gamma=1.0$ (satisfying the important scaling relation
  $\gamma=(\alpha-1)/(\tau-1)$) do not coincide with the expectations
  for an Ornstein-Uhlenbeck process.  
\label{fig:PLaws}}
\end{figure}
\par\end{center}
First we discuss computational results and then we employ scaling
arguments to explain the findings.  On the one hand, as already shown
in Fig.\ref{fig:OU} using $S$, i.e. criterion A, one reproduces the
expected theoretical results for all three avalanche exponents. On the
other hand, as illustrated in Figure \ref{fig:PLaws}, the statistics
of avalanche sizes, as determined employing $\Sigma$ for an
Ornstein-Uhlenbeck process is anomalous and does not match the
expectations for the theoretically known values i.e. the measured
value $\tau \approx 3/2$ does not coincide with the expected value
$\tau=4/3$. In particular, the numerical observation of,
$\tau \approx 3/2$ could be (wrongly) taken as evidence of branching
process-like scaling \cite{Linken2012,Larremore2014}.  The fact that
there is something suspicious with the definition (ii) can be
noticed observing that both sizes and durations scale in the same
way, entailing $\gamma=(\alpha-1)/(\tau-1)=1$, which would imply a locally
linear (i.e. ``tent like'') shape of avalanches \cite{Colaiori,shape,Laurson}.

\subsubsection{Scaling arguments}

The correction $s^*$ for a given avalanche (such that
$\Sigma= S+ s^*$) is nothing but $s^*=\theta T$, where $T$ is the
avalanche duration.  The distribution of first-passage times for the
Wiener process is given by $P(T) \sim T^{-3/2}$. {As said above,
  the same result holds for the Ornstein-Uhlenbeck case up to an upper
  cut-off.  Thus, $\Sigma$ has a correction $s^*$ with respect to $S$
that scales as the avalanche duration: $P(s^*) \sim {s^*}^{-3/2} $.
Assuming that the probability to observe a given size $S$
  \emph{conditioned} to a given avalanche duration $T$, 
$P(S|T)$, is a peaked function around its mean value (as
usually occurs for avalanches \cite{Avalanches}) and using the fact
that $\langle S \rangle \sim T^{\gamma}$, with $\gamma=3/2$, then
\begin{equation}
\Sigma(T)=S(T)+s^*(T)= \tilde{c} T^{3/2}+\theta T, \label{2terms}
\end{equation}
from where it follows that
\begin{equation}
d\Sigma=(cT^{1/2}+\theta) dT
\label{3}
\end{equation}
Thus, we can readily write (using the implicit function theorem):
\begin{equation}
P(\Sigma(T))=P(T)\frac{dT}{d\Sigma}={\cal N}\frac{T^{-3/2}}{cT^{1/2}+\theta}.
\end{equation}
From this, in the limit of vanishing threshold $\theta$ in Eq.(\ref{3}), one has 
\begin{equation}
P(\Sigma)\approx\frac{{\cal N}}{c}T^{-2}\approx\frac{{\cal
    N}}{c}[(\Sigma/c)^{2/3}]^{-2}\sim \Sigma^{-4/3},
\end{equation}
which is the correct result for the avalanche size distribution of an
Ornstein-Uhlenbeck process. On the other hand, for larger values of
the threshold $\theta$ and relatively small values of $T$ (and, thus,
also typically small values of $\Sigma$) one has
\begin{equation}
P(\Sigma)\approx\frac{{\cal N}}{\theta}T^{-3/2}\approx\frac{{\cal
    N}}{\theta}[\Sigma/\theta]^{-3/2}\sim\Sigma^{-3/2},
\end{equation}
in agreement with the numerical observation above. In other words, the
additional contribution $s^*$ dominates the scaling behavior of the
avalanche size $\Sigma$ when $\theta$ is relatively large. It is
important to emphasize that, in any case, one should recover the
correct asymptotic value --i.e. the behavior for large values of
$\Sigma$-- of the avalanche size exponent ($\tau=4/3$) for any value
of $\theta$ but this requires going to larger and larger avalanche
sizes as $\theta$ is chosen larger and larger. In particular, Figure
\ref{fig:crossover} illustrates that there is a crossover from the
value $\tau \approx 3/2$ measured for small avalanche-sizes to the
true asymptotic scaling $\tau =4/3$, for larger sizes. The crossover
point grows with $\theta$, so that the effect is not observed for
$\theta \approx 0$, but may extend for many scales even for moderate
values of $\theta$. In particular, given that an upper cut-off to
scaling may exist (controlled e.g. by $1/a$ in the case of an
Ornstein-Uhlenbeck process that we are considering here or by 
finite-size effects), the transient behavior usually extends all the
way up to the cut-off, so that the true asymptotic behavior can be
unobservable with criterion B if large values of $\theta$ are considered.

Thus, summing up, considering criterion B for the definition of
avalanche sizes together with relatively large threshold values or not
sufficiently large statistics, may lead to the observation of an
effective value $\tau \approx 3/2$; this may induce a
misinterpretation of the scaling universality class, suggesting it is
branching-process-like rather than what it actually is: a
random-walk-like process.

Let us emphasize that the previous discussion has been done for an
Ornstein-Uhlenbeck process. However, it perfectly illustrates the
problem associated with criterion B in more general circumstances,
e.g. it also applies to asymmetric processes as the one sketched in
Fig.2. In any case, criterion B mixes the scalings of actual sizes and
times, leading to potential interpretation errors in general
stochastic processes. This problem is at the origin of
miss-classification of scaling behavior in existing works analyzing
neuronal avalanches (see e.g. \cite{Linken2012,Larremore2014}).

\begin{center}
\begin{figure}[h]
\begin{centering}
\includegraphics[width=1.0\columnwidth]{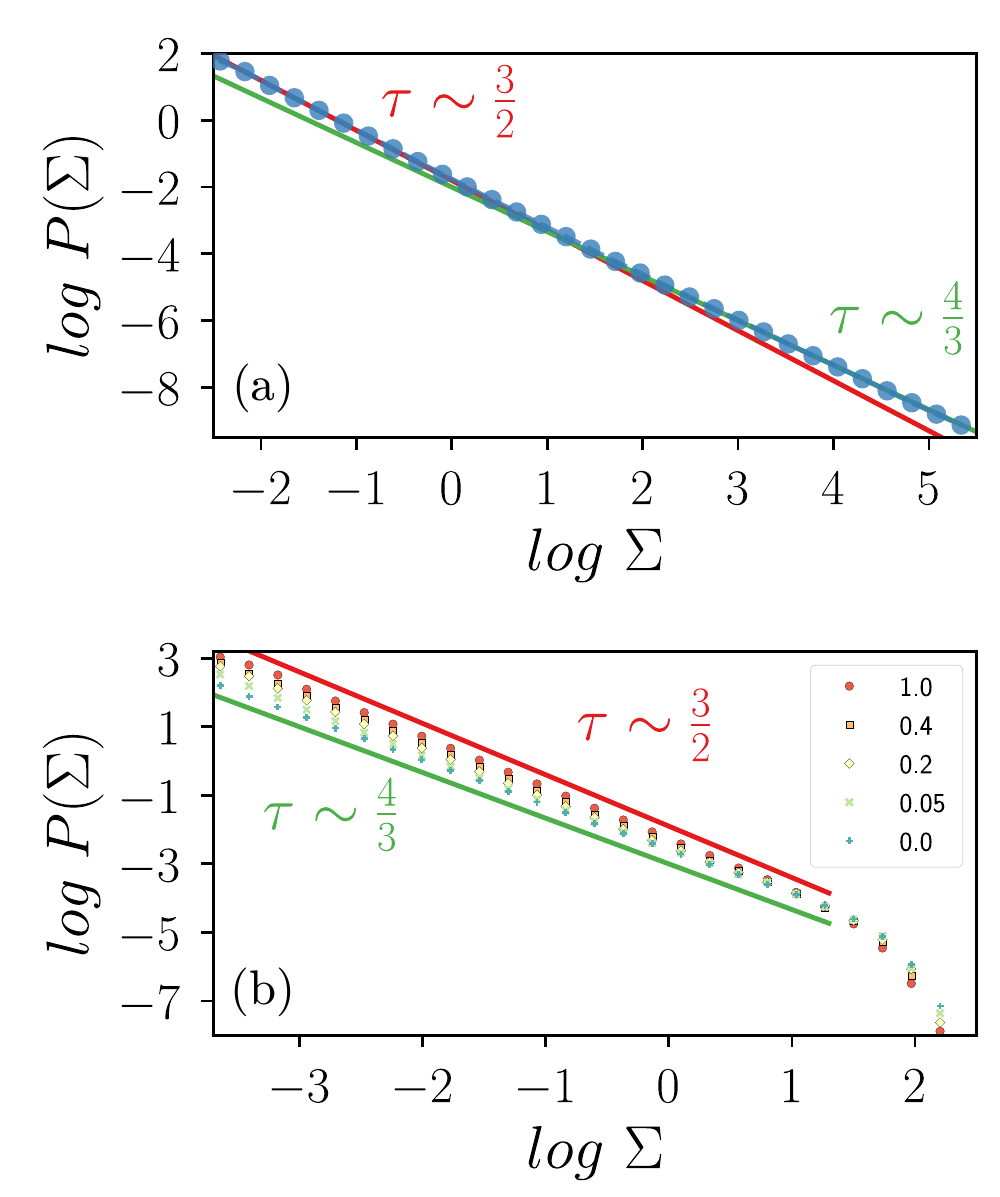}
\par\end{centering}
\caption[]{(Color online) Distribution of avalanche sizes for a
  stochastic (Ornstein-Uhlenbeck) process employing $\Sigma$ for the
  measure of avalanche sizes.  (a) Avalanche-size distribution for
  the case $\theta=1$ (with $a=0$ and $\sigma=1$ in this case):
  observe that the true exponent value $\tau=4/3$ is asymptotically
  recovered for large avalanche sizes. (b) Distribution of
  avalanche sizes for different values of the threshold parameter,
  $\theta$. The associated exponent changes continuously between the
  two limiting exponents $3/2$ (for large thresholds) and $4/3$ for
  sufficiently small ones (parameter values: $a=0.1$, $\sigma=0.5$;
  thresholds as marked in the figure legend).  
  \label{fig:crossover}}
\end{figure}
\par\end{center}

\section{Conclusion}
\newb{
In this brief paper we have shown that an inappropriate definition of
avalanche sizes as measured as excursions above a given threshold
  in continuous timeseries can lead to misleading conclusions.}
To illustrate this, we have studied a simple Ornstein-Uhlenbeck
  process (representing e.g. the time course of activity in a
mesoscopic model of neural activity) and have measured avalanches
sizes in two possible ways: (i) as the integrated activity $S$ over a
given threshold and (ii) integrating the total activity signal in
between two threshold crossings, as illustrated in
Fig. \ref{fig:poil}.  We have shown both computationally and using
scaling arguments that this latest definition can induce strong biases
in the determination of the avalanche-size exponent $\tau$.

In particular, if large values of the threshold $\theta$ are
considered, then --for relatively small avalanches-- one observes the
exponent value $\tau \approx 3/2$ which could lead to the erroneous
interpretation that an effective un-biased branching process dynamics
exists. On the other hand, for sufficiently small threshold values and
for sufficiently large avalanche sizes the correct scaling $\tau=4/3$
is recovered.  As discussed above the problem associated with
criterion B extends to any type of stochastic process as it mixes up
the scaling of actual sizes with that of durations, giving rise to
misleading results.

This is the underlying reason why recent analyses of avalanches in
mesoscopic models of neural activity that consider relatively large
thresholds \cite{Linken2012,Larremore2014,Dalla2018})
obtain $\tau\approx 3/2$ --compatible with the scaling of a critical
branching process--, a result that our analysis reveal that is not
asymptotic. Their corresponding underlying dynamics describe
fluctuations around a given mean value and, thus, the associated
avalanches should be related to excursions of random walkers and not
to critical branching processes.

As an important final remark, let us stress that it is essential
  --and it should always be done-- to consider the full set of
  avalanche exponents i.e. $\tau$, $\alpha$, and $\gamma$ as well as
  the scaling relations between them, in order to avoid possible
  errors and misleading interpretations and properly identify the type
  of scaling behavior.

\vspace{1.5cm}

\begin{acknowledgments}

  We acknowledge the Spanish Ministry and Agencia Estatal de
  investigaci{\'o}n (AEI) through grant FIS2017-84256-P (FEDER funds)
  for financial support.  This study has been partially financed by
  the Consejería de Conocimiento, Investigación y Universidad, Junta
  de Andalucía and European Regional Development Fund (ERDF),
  ref. SOMM17/6105/UGR as well as to TEACH IN PARMA for support to
  M.A. M. We thank V. Buend{\'\i}a for very useful comments and a
  critical reading of the manuscript.

\end{acknowledgments}

%

\end{document}